\documentstyle[preprint,aps,psfig]{revtex}
\tighten
\begin{document}
\draft
\preprint{\vbox{Submitted to Physical Review C
                \hfill FSU-SCRI-99-43 \\}}
\title{Strange matter in the string-flip model}
\author{Danielle Morel and J. Piekarewicz}
\address{Department of Physics and Supercomputer 
         Computations Research Institute, \\ 
         Florida State University, 
         Tallahassee, FL 32306, USA}
\date{\today}
\maketitle

\begin{abstract}
We employ variational Monte Carlo methods to study the
transition to strange matter in a simple one-dimensional 
string-flip model with two flavors and two colors of quarks. 
The dynamics of the system are described in terms of a 
many-body potential that confines quarks within hadrons, 
yet enables the hadrons to separate without generating 
unphysical long-range van der Waals forces. The model 
has ``natural'' low- and high-density limits: it behaves 
as a system of isolated hadrons at low density and as a 
Fermi gas of quarks at high density. We show that the 
system exhibits a transition to strange matter characterized 
by an increase in the length-scale for confinement. Yet 
the small increase at the transition region --- of only ten
percent --- suggests that clustering correlations remain 
strong well into the strange-matter domain. Our results 
put into question descriptions of strange matter in terms 
of noninteracting, or weakly interacting, quarks.
\end{abstract}

\section{\bf INTRODUCTION}
The possible existence of stable strange matter --- more 
stable even than ${}^{56}$Fe --- is fascinating 
indeed~\cite{review98}. That this scenario is plausible at high 
baryon density is simple to understand: for two-flavor (up and 
down) matter the Fermi energy becomes so large that the addition 
of strange quarks becomes energetically favorable, in spite of 
the larger strange-quark mass. The existence of strange
matter should have a dramatic impact on fields ranging 
from astrophysics and cosmology to nuclear and particle physics. 
Indeed, it is widely accepted that the high-density environment 
at the core of neutron stars constitutes a fruitful ground for 
the formation of quark matter. Moreover, the recent proposal of 
a ``spin-up'' phase in the timing structure of pulsars, speculated 
to signal the transition from hadronic to quark matter, makes the 
detection of this exotic phase of matter within observational 
reach~\cite{GPW97}. These astronomical observations complement 
terrestrial searches for strange matter which have improved 
considerably, and will continue to do so, with the commission of 
new powerful relativistic heavy-ion colliders~\cite{Arms99}.

The theoretical study of strange matter has traditionally been
conducted in two different pictures: one using a hadronic 
model~\cite{Dover93} --- similar to ordinary nuclei --- where 
the fundamental degrees of freedom are mesons and baryons, another
using a quark 
model~\cite{Witten84,FarJaf84} consisting of 
massless, noninteracting quarks confined inside a bag (note that
the inclusion of a finite strange-quark mass and a perturbative 
treatment of QCD does not seem to change the qualitative picture).  
Presumably a description of strange matter in terms of mesons and 
baryons is well motivated in the low-density regime where clustering 
correlations remain important. Alternatively, strange matter viewed
in terms of a Fermi gas of quarks might be appropriate in the 
high-density domain, as now the average interquark separation 
becomes considerably smaller than the length-scale for confinement.  
This division, however, seems arbitrary. Indeed, it is not at all 
clear how and at what density the ``transition'' should be made 
from a hadronic- to a quark-matter description. Moreover, there 
might be certain regions where none of the pictures could be 
physically realized. For example, one could imagine the existence 
of an intermediate-density phase characterized by an increase in 
the length-scale for confinement (``swollen'' baryons) but also 
in which clustering correlations remain important. Perhaps the 
most serious difficulty encountered in studying the density 
dependence of strange matter is how to model a system that
has quarks confined inside color-neutral hadrons at low density but
free quarks at high density. The divorce of the two pictures is
further reinforced by the difficulty of treating quark confinement: 
how can quarks be confined inside hadrons, yet the hadrons separate
without generating long-range van der Waals forces? We offer no new
insight into this difficult problem; rather, we argue here that it 
is useful to consider an effective model which interpolates between 
a hadron- and a quark-based description.

To do so we rely on the string-flip model. One strength of this model 
is that the evolution of the system from a low-density hadronic
phase to a high-density quark phase is dynamical rather than through
the introduction of artificial parameters. Yet the string-flip model
employed here is simple and its predictions should be taken with
caution. Still, we trust that some of the qualitative features
obtained from our analysis will have a direct correspondence to
those made in more sophisticated models. The string-flip model
used here is an offspring of the original one-dimensional model
proposed by Lenz and collaborators~\cite{Lenz86} and applied for the
first time to nuclear matter by Horowitz, Moniz, and
Negele~\cite{HMN85}.  Since then the model has been generalized to
three dimensions and internal degrees of freedom (such as color) have
been incorporated to study a variety of ground state
properties~\cite{HorPan85,Watson89,HorPie92,FriPie94}. All these
models represent simple --- yet realistic --- many-body generalizations
of the non-relativistic constituent quark model~\cite{IsgKar79}. The
cornerstone of the string-flip model is the many-body nature of the
potential which allows quark confinement within hadrons, while
enabling the hadrons to separate without generating long-range van der
Waals forces~\cite{GreLip81}.

There are several questions that we will try to answer in this paper.
First, how does the length-scale for quark confinement evolve as the
density of the system increases and how is this evolution affected by
the presence of strange quarks? Second, how does the many-body
(confining) potential modifies the transition to strange matter
relative to the predictions of a Fermi-gas estimate; do clustering
correlations remain important at the transition density or has the
system dissolved into a collection of noninteracting quarks? Finally,
what can one learn about the dynamic response of the system through a
study of two-body correlations between quarks.

Our paper has been organized as follows. In Sec.~\ref{sec:formal} we
introduce the particular version of the string-flip model used here to
compute a variety of ground-state properties of strange matter.
Variational Monte-Carlo results are presented in
Sec.~\ref{sec:results} and compared against some useful limiting
cases. Finally, our conclusions and outlook are offered in 
Sec.~\ref{sec:conclusions}.

\section{Formalism}
\label{sec:formal}

We compute the properties of strange matter in a string-flip model
that has constituent quarks --- interacting through a many-body
potential --- as the fundamental degrees of freedom. The many-body
potential used here is explicitly symmetric in all quark coordinates,
confines quarks within hadrons, and enables the hadrons to separate
without generating long-range van der Waals forces. In this paper we
use the simplest form of this model that can describe the basic quark
structure of strange matter. In this version quarks are constrained 
to move in one spatial dimension and have only two SU(2)-like 
intrinsic degrees of freedom: color and flavor. The color --- a 
global, rather than a local --- quantum number can be either red or 
blue (with red+blue=``white'')  and the flavor either up or strange. 
Thus, ``color-neutral hadrons'' are represented as two-quark 
composites coming in three different varieties: nucleon ($uu$), 
lambda ($us$), and cascade ($ss$). 
 
\subsection{Many-quark Potential}
\label{subsec:MQP}

The Hamiltonian for a system of $N$ quarks moving in one-spatial 
dimension under the influence of the many-body potential $V$ is 
given by
\begin{equation}
   H=T+V=\sum\limits^N_{i=1} \frac {p^2_i}{2m_i} + V(x) \;.
 \label{hamil}
\end{equation}
The first step in constructing the $N$-body potential $V(x)$ is 
the pairing of all $N$-quarks into $A\!=\!N/2$ color-neutral 
clusters, irrespective of quark flavor. This procedure results 
in a total of $A!$ possible configurations. For each of these
configurations one now introduces a flux-tube energy:
\begin{equation}
   V_{n}(x) = \frac{1}{2}k 
   \sum\limits^A_{i=1}\left(r_i^{(n)} - b_i^{(n)}\right)^2 \;,
 \label{fluxtube}
\end{equation}
where $r_i^{(n)}$ and $b_i^{(n)}$ are the positions of the red 
and blue quarks within the $i$th cluster, and $k$ is the strength 
of the confining potential (which here has been assumed harmonic). 
Finally, the many-body potential is obtained by demanding that 
the pairing of quarks into color-neutral hadrons be optimal. 
That is, that the potential be the minimum value of $V_n$ among 
all the $A!$ possible pairings,
\begin {equation}
    V(x) \equiv \min_{n=1}^{A!} \Big\{V_{n}(x)\Big\}\;.
 \label{vopt}
\end{equation}
Note that in what follows we adopt units in which $\hbar$, the 
spring constant $k$, and the up-quark mass are all equal to one while 
for the constituent strange-quark mass the canonical value of 
$m_{s}/m_{u}=1.6$ is used. 

The minimum prescription described above ensures that the potential is,
indeed, symmetric under the exchange of identical quarks.  Moreover,
no long-range van der Waals forces are generated in the model as the
force saturates within each individual hadron. In this way 
residual interactions between hadrons involve the possibility of
quark-exchange (when the strings flip) and the Pauli exclusion
principle between identical quarks. Finally, as it is possible for all
pairings to change as one moves a single quark, the potential is truly
an $N$-body operator; it cannot be reduced to a conventional sum of
two-body terms. The global minimization procedure should be taken
among all possible $A!$ configurations, which rapidly becomes
unmanageable; for $A\!=\!40$ (the canonical value used in our
simulations) the total number of possible configurations is close to
$10^{48}$. Fortunately, because we have restricted the problem to one
spatial dimension, finding the optimal solution scales linearly ---
not exponentially --- with the number of quarks. (In reality the
solution to the pairing problem scales only with the cube of the
number of quarks~\cite{HorPie92,BurDer80}. To our knowledge a
polynomial solution to the three-quark assignment problem has yet to
be found).

To generate the desired $A$ configurations one starts by ordering the 
red and blue quarks, independently, in a one-dimensional box of length 
$L$ (note that the concept of ``order'' is meaningful only in one 
spatial dimension). The first of these configurations is obtained by
mapping the first red quark into the first blue quark, the second red 
quark into the second blue quark, and so on until finally the last red 
quark is paired with the last blue quark. Similarly, the second of these 
configurations is generated by mapping the first red quark into the 
second blue quark and then continuing the assignments sequentially. One 
continues this procedure until in the last configuration the first 
red quark is mapped into the last blue quark. That is,
\begin{eqnarray*}
   r_1   \leftrightarrow b_1   \;; \; 
   r_2   \leftrightarrow b_2   \;; \; &\ldots& \;; \; 
   r_{A} \leftrightarrow b_{A} \;,
   \quad\mbox{or}\quad 
   V_{1}(x) = \frac{1}{2} 
   \sum\limits^A_{i=1} \left(r_i - b_i\right)^2 \;, \\
   r_1   \leftrightarrow b_2   \;; \; 
   r_2   \leftrightarrow b_3   \;; \; &\ldots& \;; \; 
   r_{A} \leftrightarrow b_{1} \;,
   \quad\mbox{or}\quad
   V_{2}(x) = \frac{1}{2} 
   \sum\limits^A_{i=1} \left(r_i - b_{i+1}\right)^2 \;, \\
        &         &                     \ldots          \\
   r_1   \leftrightarrow b_A   \;; \; 
   r_2   \leftrightarrow b_1   \;; \; &\ldots& \;; \; 
   r_{A} \leftrightarrow b_{A-1} \;,
   \quad\mbox{or}\quad 
   V_{A}(x) = \frac{1}{2} 
   \sum\limits^A_{i=1} \left(r_i - b_{i+A-1}\right)^2 \;.
\end{eqnarray*}
The potential energy is then selected to be the minimum among the
$A$ configurations: 
\begin {equation}
    V(x) = \min 
   \Big\{V_{1}(x),V_{2}(x),\ldots,V_{A}(x)\Big\}\;.
 \label{vopt1D}
\end{equation}
Despite the relative simplicity of the model a clear picture of the
quark dynamics emerges in the low- and high-density phases of the
system. In the low-density phase the confinement scale --- a typical
distance between quarks within a hadron in free space  --- is much 
smaller than the inter-hadron separation. Since the potential
saturates within each individual hadron, quark exchange is suppressed 
and the system resembles a collection of noninteracting nucleons.
In the high-density phase, however, the confinement scale becomes 
larger than the inter-hadron separation making  
string rearrangement dominant. By then the potential energy 
is unimportant relative to the kinetic energy and the system 
resembles a free Fermi gas of quarks.

\subsection{Fermi Gas of Quarks}
\label{subsec:FFG}

In this section we compute ground-state properties of a simple 
Fermi-gas model of quarks in order to establish a baseline against 
which the predictions of the string-flip model may be compared. The 
total energy of a one-dimensional free Fermi gas of quarks is 
simply the sum of the kinetic energies of the two flavors
\begin{equation}
  \frac{T_{FG}}{N}(\rho,\sigma)                            =
     (1-\sigma)+\frac{k_{F}^{2}}{6} (1-\sigma)^{3}    +
   m_{s}\sigma +\frac{k_{F}^{2}}{6m_{s}}\sigma^{3}    \;;\quad
   k_{F}\equiv\frac{\pi}{2}\rho                       \;,
 \label{tfg}
\end{equation}
where the strangeness-per-quark parameter $\sigma\equiv|S|/N$ 
and the density $\rho$ have been introduced. However our results 
will be presented in terms of the more conventional strangeness-per-baryon 
ratio $f_{s}=|S|/A$ and the relative density $\rho/\rho_{c}$ where 
$\rho_{c}$ is defined below. Note that $k_{F}$ is the Fermi momentum 
of the system in the limit of $|S|\equiv 0$. The strangeness-per-quark 
ratio is determined by demanding that the total energy of the system 
be minimized with respect to $\sigma$ at fixed baryon density. That is,
\begin{equation}
  \left(\frac{\partial T_{FG}}
  {\partial\sigma}\right)_{\rho}\!=0=
  -1-\frac{k_{F}^{2}}{2}(1-\sigma)^{2} +
  m_{s}+\frac{k_{F}^{2}}{2m_{s}}\sigma^{2} \;.
 \label{mu}
\end{equation}
The above condition represents chemical equilibrium meaning 
that the chemical potential of both species must be equal. The 
solution to this quadratic equation is simple and from it one 
obtains the strangeness-per-quark ratio as a function of the density 
of the system
\begin{equation}
 \sigma(\rho)=\frac{m_{s}k_{F}-
              \sqrt{m_{s}\Big[2(m_{s}-1)^{2}+k_{F}^{2}\Big]}}
              {(m_{s}-1)k_{F}} \;.
 \label{sigma}
\end{equation}
Although not explicitly stated, the above equation is valid only
above a critical density $\rho_{c}$ below which the 
strangeness-per-quark ratio vanishes. This is due to the fact that 
below $\rho_{c}$ it is not yet energetically favorable to 
introduce strange quarks into the system. The critical density 
is given by
\begin{equation}
   \rho_{c}=
   \frac{2}{\pi}\sqrt{2(m_{s}-1)}\approx0.697 \;.
 \label{rhoc}
\end{equation}
Finally, the Fermi-gas energy is obtained by substituting
Eq.~(\ref{sigma}) into Eq.~(\ref{tfg}), i.e.,
\begin{equation}
  T_{FG}(\rho)=T_{FG}\Big(\rho,\sigma(\rho)\Big) \;.
 \label{tfgrho}
\end{equation}

Another observable useful in characterizing the transition 
from hadronic- to quark-matter is the two-body correlation 
function. This function represents the ground-state probability 
of finding a pair of quarks at a fixed separation. That is,
\begin{equation}
  \rho_{2}(r)=\left\langle\sum_{i<j}
	      \delta\Big(r-|r_{i}-r_{j}|\Big)
              \right\rangle =
	      \sum_{q}e^{iqr}\sum_{k_1,s_1;k_2,s_2}
              \left\langle C^{\dagger}_{k_2+q,s_2}
                           C^{\dagger}_{k_1-q,s_1}
              C_{k_1,s_1}C_{k_2,s_2}\right\rangle \;,
 \label{rho2}
\end{equation}
where $C^{\dagger}_{k,s}$($C_{k,s}$) represents a quark 
creation(annihilation) operator with momentum $k$
and (SU(2)-like) color quantum number $s$. The above
two-body operator is easily evaluated in the Fermi-gas
limit to obtain
\begin{equation}
  \rho_{2}(r) \propto g_{FG}(r) = 1-\frac{1}{2}
  \left(\frac{\sin(k_{F}r)}{k_{F}r}\right)^{2} \;.
 \label{rho2fg}
\end{equation}
Note that we have introduced the function $g_{FG}(r)$
which is simply the two-body correlation function 
normalized to one at large distances. This function
is sensitive only to Pauli correlations.  The second 
term in Eq.~(\ref{rho2fg}) --- with its corresponding 
50\% reduction at the origin --- accounts for this fact.

\subsection{Variational Wavefunction}
\label{subsec:VW}

In this section we compute ground-state properties of the
string-flip Hamiltonian of Eq.~(\ref{hamil}) using a variational 
Monte Carlo approach. For a system of $N$ quarks moving in a 
one-dimensional box of length $L$ the one parameter variational
wavefunction has the form
\begin{equation}
  \Psi_{\lambda}=e^{-\lambda U(x)} \Psi_{FG}(x) \;,
 \label{psivar}
\end{equation}
where $\lambda$ is the variational parameter and $\Psi_{FG}$ 
represents a free Fermi-gas wavefunction comprised of the 
product of four Slater determinants; one for each color-flavor 
combination. For example, the one-dimensional Slater determinant
for the ``red-up'' combination is of the form~\cite{HMN85,HorPie91}
\begin{equation}
  \Phi_{\rm FG}^{(red-up)}(x) =
  \prod\limits_{i<j}\sin\left[\frac{\pi}{L}
               (r^{up}_i -r^{up}_j) \right] \;.
 \label{phifg}
\end{equation}
The Fermi-gas wavefunction is exact for a system of fermions with 
no correlations other than those generated by the Pauli exclusion 
principle, while the exponential portion of the variational 
wavefunction --- symmetric under the exchange of identical quarks 
--- characterizes the amount of clustering in the ground state. 
Here we have introduced the ``quasi-potential''
\begin{equation}
   U(x) = \alpha_{uu} V_{uu}(x) 
        + \alpha_{us} V_{us}(x) 
        + \alpha_{ss} V_{ss}(x) \;,
 \label{uofx}
\end{equation}
where $V_{uu}$, $V_{us}$, and $V_{ss}$ are the portions of the
many-body potential that reside in nucleons, lambdas, and
cascades respectively; that is: $V(x)=V_{uu}(x)+V_{us}(x)+
V_{ss}(x)$. The constants $\alpha_{uu}$, $\alpha_{us}$, and 
$\alpha_{ss}$ correct for the sizes of the different hadrons 
and are thus proportional to the ratio of the reduced mass 
of the corresponding hadron to that of the nucleon:
\begin{equation}
   \alpha_{uu} = 1\;, \quad
   \alpha_{us} = \sqrt{\frac{2m_{s}}{1+m_{s}}} \approx 1.109 \;, 
     \quad\mbox{and}\quad
   \alpha_{ss} = \sqrt{m_{s}} \approx 1.265 \;. 
\end{equation}
In the low-density phase --- or in a phase with various noninteracting 
hadrons --- the variational wavefunction becomes exact in the limit 
of $\lambda\rightarrow 1/\sqrt2$ (value of the oscillator parameter 
for isolated two-quark clusters). Similarly, in the high-density limit 
where the potential is unimportant, the variational wavefunction 
reproduces exactly that of a free Fermi gas of quarks in the limit 
of $\lambda \rightarrow 0$.

\subsection{Metropolis Monte Carlo}
\label{subsec:MMC}

An additional advantage of using such a simple variational
wavefunction is that the expectation value of the kinetic and 
potential energies are not independent~\cite{HMN85,HorPie91}.
Indeed, the expectation value of the kinetic energy, difficult 
to simulate because of its derivatives, is easily related to 
the expectation value of the potential energy. That is,
\begin{equation}
  \langle\Psi_{\lambda}|T|\Psi_{\lambda}\rangle = 
     T_{FG} + 2 \lambda^2 
  \langle\Psi_{\lambda}|V|\Psi_{\lambda}\rangle \;,
 \label{virial}
\end{equation}
where $T_{FG}$ is the kinetic energy of a one-dimensional free Fermi
gas given in Eq.~(\ref{tfg}). The minimum value of the kinetic energy
is attained in the Fermi-gas limit.  The additional term in the above
expression --- $2\lambda^2\langle V \rangle_{\lambda}$ --- represents
the increase in the kinetic energy above the Fermi-gas limit due to
the presence of clustering correlations. While the kinetic energy
disfavors such correlations the potential energy favors them, at
least at low density. The dynamic competition between the kinetic 
and potential energies will yield an optimal value for the variational 
parameter for any given density and strangeness-to-baryon ratio. The 
optimal value of the variational parameter is thus obtained by 
minimizing the total energy of the system:
\begin{equation}
  \langle E \rangle_{\lambda\rho\sigma} = 
  T_{FG}(\rho,\sigma) + 
  (2\lambda^2+1) \langle V \rangle_{\lambda\rho\sigma} \;.
 \label{evar}
\end{equation}
As promised, to compute the ground-state expectation value of the 
energy we need only compute the expectation value of the 
potential energy, which we accomplish using Monte-Carlo methods. 
Using the algorithm of Metropolis {\it et al.,} it can be
shown~\cite{NegOrl88} that the expectation value of the potential 
energy can be obtained by performing a suitable average of the form
\begin{equation}
   \langle V \rangle = \lim_{M\rightarrow\infty}
   \frac{1}{M} \sum\limits_{m=1}^M V(x_{m}) \;,
 \label{metro}
\end{equation}
where the sequence of points $\{x_{m}\}$ are distributed according
to the square of the variational wavefunction. It then becomes a 
simple --- yet computationally-intensive -- task to extract the 
optimal variational parameter that minimizes the energy of the system 
for each value of the density and for each number of strange quarks.

\section{Results}
\label{sec:results}
\begin{figure}
\bigskip
\centerline{
  \psfig{figure=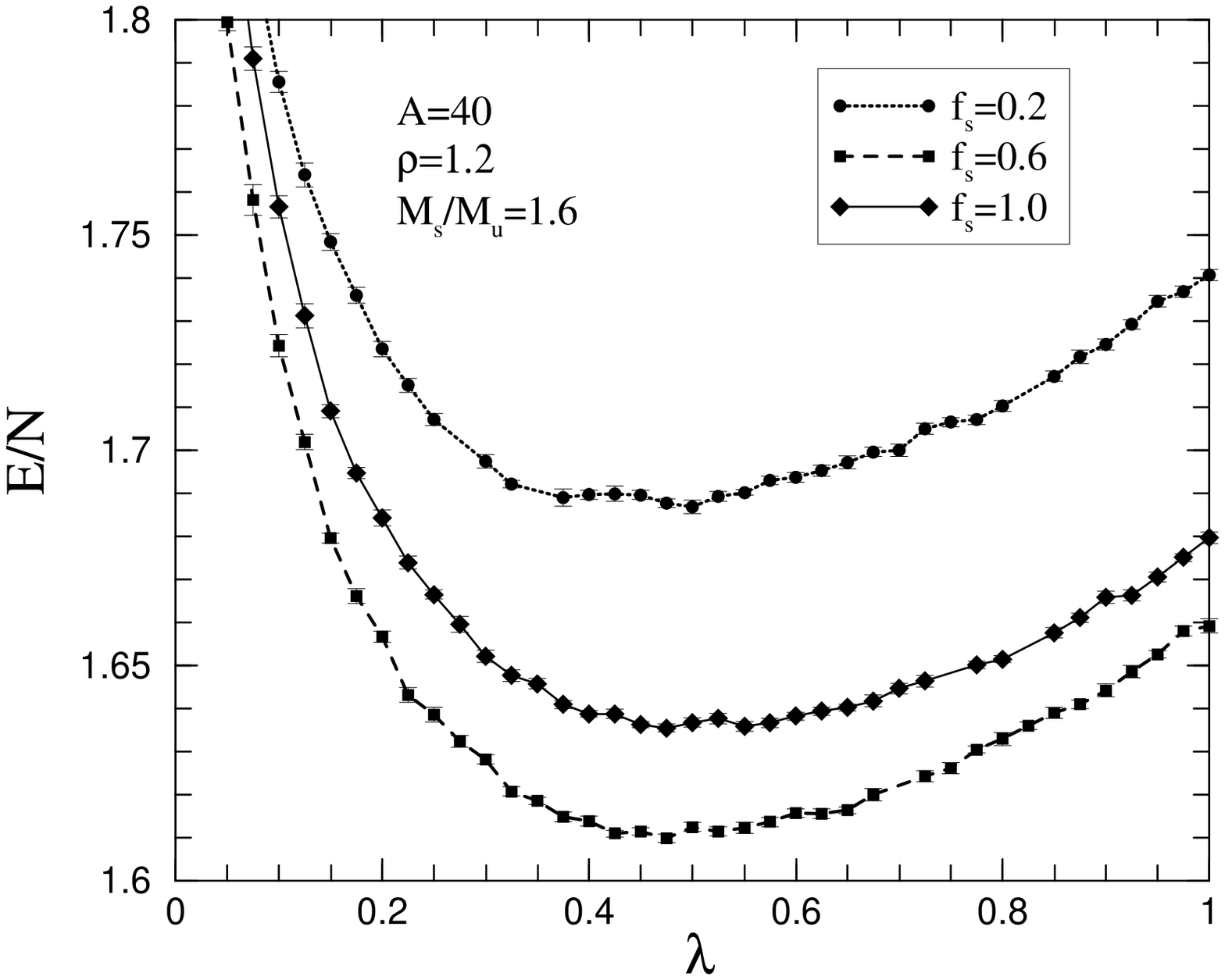,height=5in,width=5.5in,angle=0}}
 \vskip 0.1in
 \caption{Energy-per-quark as a function of the variational
          parameter for a variety of strangenes-per-baryon 
          ratios. The number of baryons has been fixed at 
          $A\!=\!40$ and the quark density at $\rho=1.2$
          ($\rho/\rho_{c}\!=\!1.72$).}
 \label{fig1}
\end{figure}

We start this section by displaying in Fig.~\ref{fig1} the
energy-per-quark of the system as a function of the variational
parameter for a variety of strangeness-per-baryon ratios at a fixed
density of $\rho=1.2$ (or $\rho/\rho_{c}=1.72$; recall that
$\rho_{c}=0.697$ represents the critical density for the transition to
strange matter in a Fermi-gas model). Note that all the calculations
reported in this work have been effected at the fixed baryon number of
$A\!=\!40$. The plot shows the, time-consuming, procedure used to
extract the optimal variational parameter, the strangeness-per-baryon
ratio, and the energy-per-quark at a fixed value of the
density. Indeed, for every value of $f_{s}$ the optimal variational
parameter is extracted as the point at which the derivative of the
energy with respect to $\lambda$ vanishes. For example, Fig.~\ref{fig1}
shows that, for this density, the minimum energy-per-quark decreases 
when the number of strange quarks goes from 8 ($f_{s}=0.2$) to 24 
($f_{s}=0.6$) but increases again if more strange quarks ($f_{s}=1.0$) 
are added to the system. The optimal value of $\lambda$ at a density 
of 1.2 must therefore be extracted when the system contains 24 quarks. 
One then repeats this procedure for all densities.

\begin{figure}
\bigskip
\centerline{
  \psfig{figure=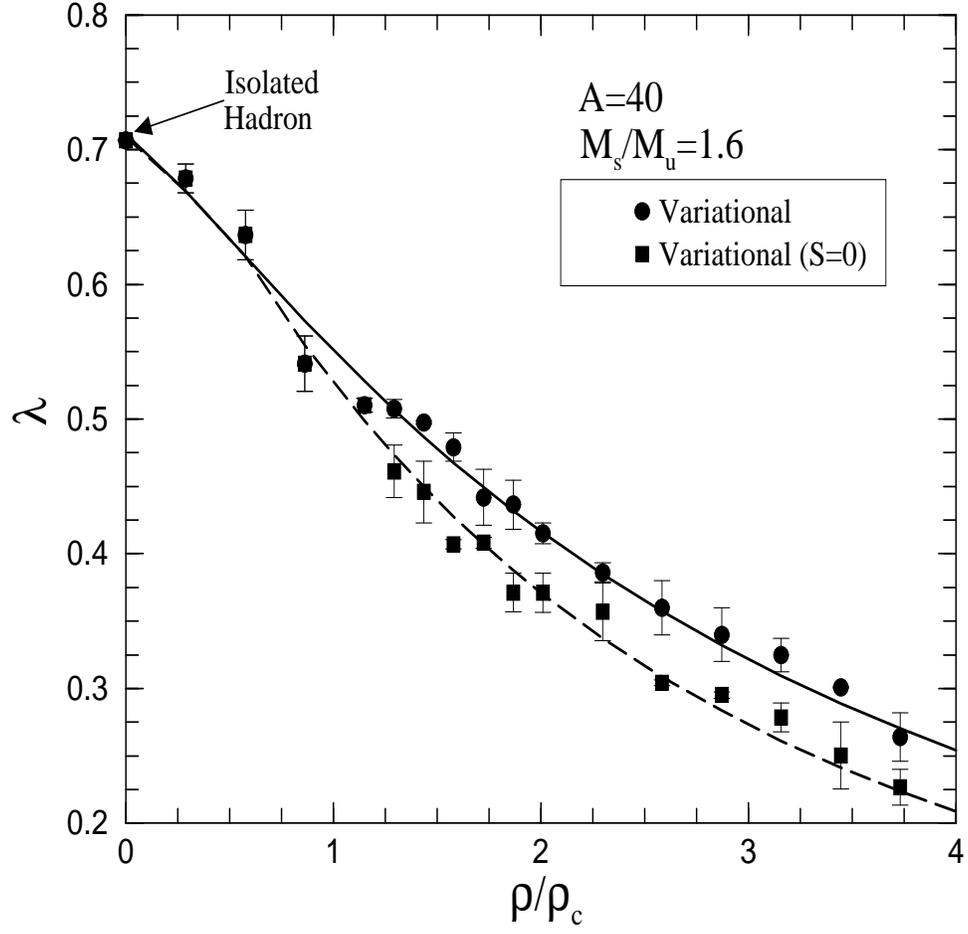,height=5in,width=5in,angle=0}}
 \vskip 0.1in
 \caption{Variational parameter as a function of density for
          nuclear (squares) and strange (circles) matter.}
 \label{fig2}
\end{figure}

The optimal variational parameter as a function of density is
displayed in Fig.~\ref{fig2}. Variational results for strange
(circles) and non-strange nuclear (squares) matter are shown; the solid
and dashed lines are smooth interpolations to the data.  As expected,
the variational parameter evolves from the isolated-hadron limit of
$\lambda_{0}\!=\!1/\sqrt{2}$ at low density towards the Fermi-gas
limit of $\lambda=0$ at high density.  It is important to stress that
the decrease in $\lambda$ with density --- with the corresponding
increase in the length-scale for confinement --- is a prediction of
the model. At the transition density to strange matter, $\rho=0.9$ or
$\rho\!=\!1.3\rho_{c}$, both sets of results start to differ, indeed 
strange matter favors stronger clustering correlations because of the 
larger mass of the strange quark. In this way we obtain one of the most
important results of our simulations: at the transition density the
length-scale for confinement (proportional to $\lambda^{-1/2}$) has
increased by only 10\%, relative to its isolated-hadron value. At 
this density --- and well into the strange-matter region ---
clustering correlations remain strong and the system is far
from becoming a collection of noninteracting, or weakly interacting,
quarks.

\begin{figure}
\bigskip
\centerline{
  \psfig{figure=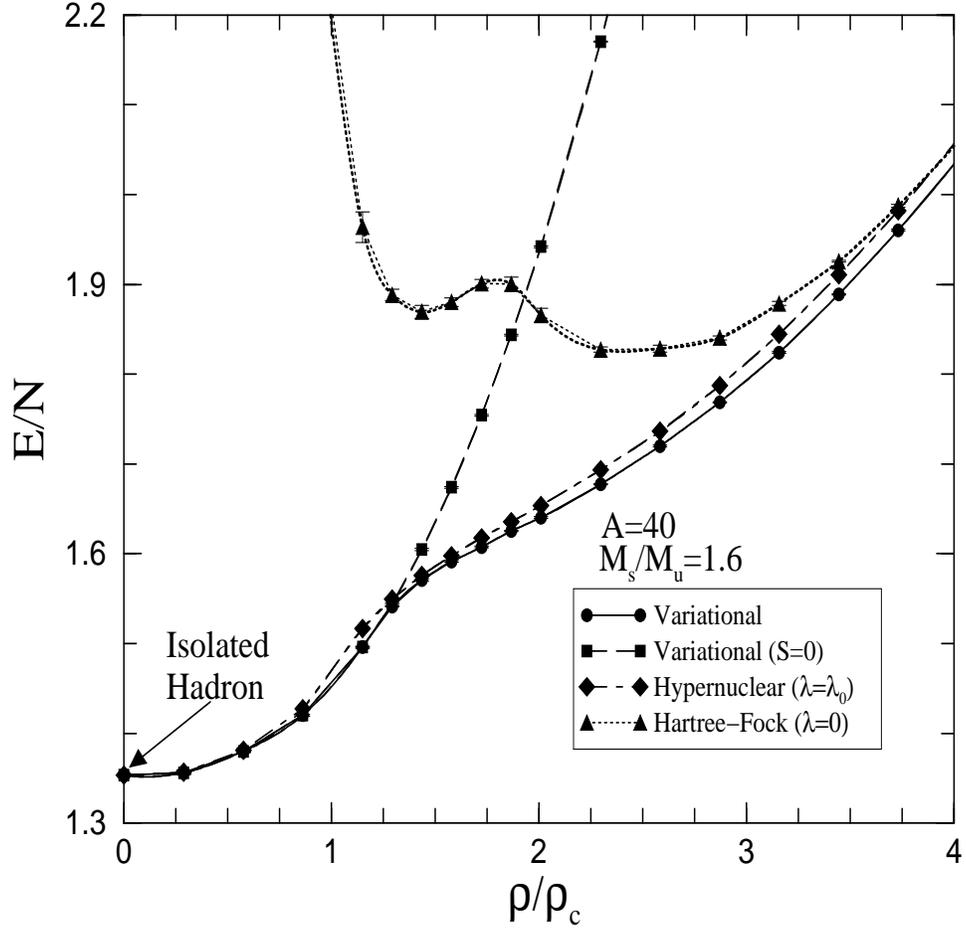,height=5in,width=5in,angle=0}}
 \vskip 0.1in
 \caption{Energy-per-quark as a function of density. 
	  See the text for details.}

 \label{fig3}
\end{figure}

The equation of state of the system --- the energy-per-quark as a
function of density --- is displayed in Fig.~\ref{fig3}. The
variational results (circles joined by the solid line) evolve
from the isolated-hadron to the Hartree-Fock limit. The importance 
of clustering correlations at low density is
irrefutable. Indeed, a Hartree-Fock estimate of the ground-state
energy (triangles joined by the dotted line) diverges at low densities
as $E_{HF}\!\sim\!\rho^{-3/2}$. Moreover, although clustering
correlations will eventually become unimportant, the energy of the
system remains about 5\% larger than the Fermi-gas result even at the
highest shown density. This picture also emphasizes that at the
transition region, depicted by the point at which the nuclear
calculation (squares joined by the dashed line) departs from the
strange-matter calculation, clustering correlations remain strong.
Finally, we have included ``hypernuclear'' results (diamonds joined by
dot-dashed lines) obtained by maintaining the variational parameter
fixed at its free-space value ($\lambda\!=\!\lambda_{0}$ at all
densities). In this approximation baryons are treated as 
incompressible quark composites. It can be seen that for the range 
of densities probed in our simulations, baryon ``swelling'' --- the 
increase in the length scale for confinement with density --- seems 
to play an insignificant role, making the hypernuclear limit an excellent
approximation to the variational results. 

\begin{figure}
\bigskip
\centerline{
  \psfig{figure=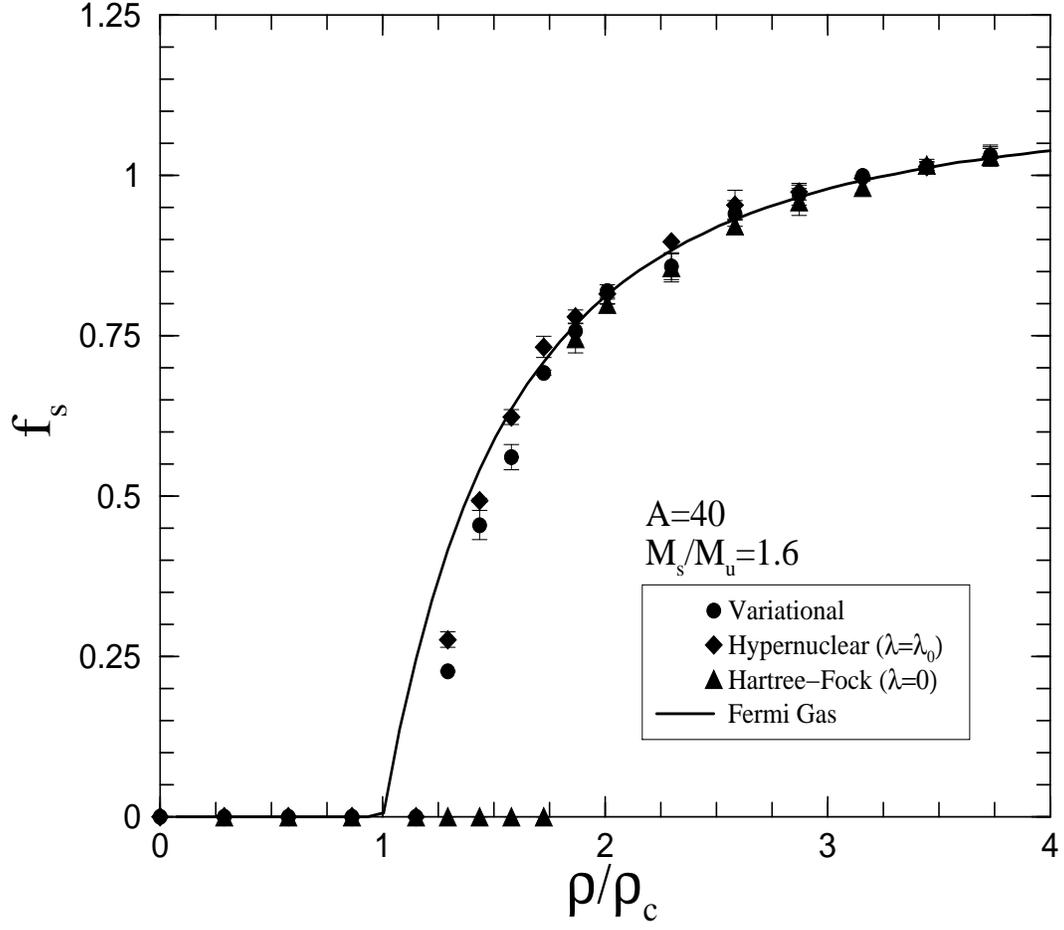,height=5in,width=5.5in,angle=0}}
 \vskip 0.1in
 \caption{Strangeness-per-baryon ratio as a function of  
	  density. See the text for details.}
 \label{fig4}
\end{figure}

Even more insensitive to baryon swelling is the strangeness-per-baryon
ratio displayed in Fig.~\ref{fig4}. Indeed, aside from a minor delay
in the transition to strange matter relative to the Fermi-gas
estimate, the strangeness-per-baryon ratio seems insensitive to most
details of the many-body potential. Clustering is the only feature
that seems to affect $f_s$. Yet even the Hartree-Fock estimate agrees 
with the variational result for $\rho\!>\!1.8\,\rho_{c}$ despite the
fact that at this density the Hartree-Fock result overestimates the 
variational energy by almost 20\%.

\begin{figure}
\bigskip
\centerline{
  \psfig{figure=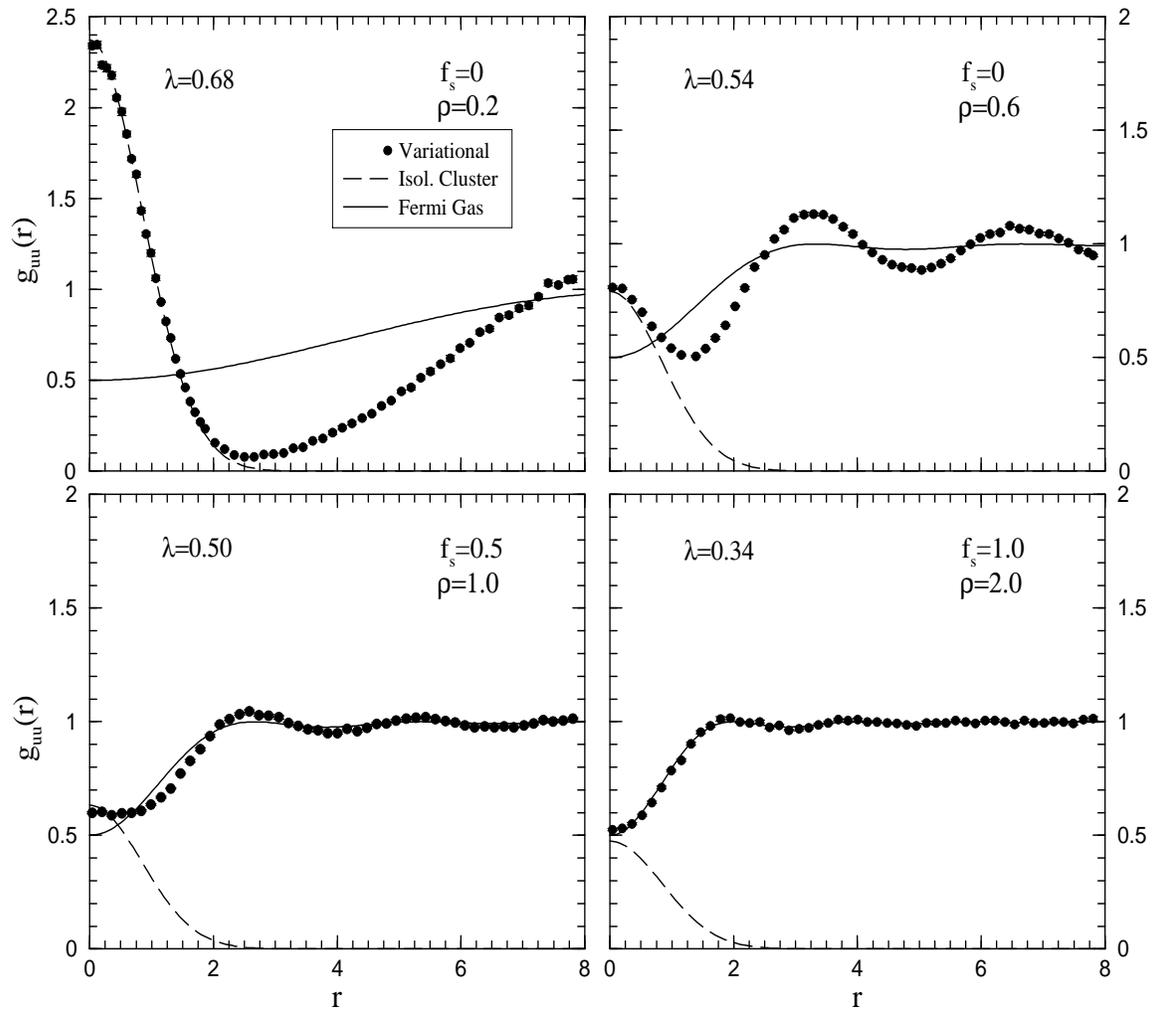,height=5.5in,width=6in,angle=-90}}
 \vskip 0.1in
 \caption{Two-body up-up correlation function for a variety 
	  of densities.}   
 \label{fig5}
\end{figure}

We continue by presenting results for the two-body
correlation function. While providing a useful characterization 
of the transition from hadronic- to quark-matter, this observable
also appears to be more sensitive to the fine details of the model. In
Fig.~\ref{fig5} we display the evolution of the ``up-up'' correlation
function with density. The upper-left-hand panel shows the system at
very low density where clustering correlations remain strong. Thus,
the two-body correlation function at short-to-intermediate distances
is accurately represented by the Gaussian behavior of the wavefunction 
of an isolated nucleon (depicted by the dashed line and barely visible in
the figure). Also shown is the two-body correlation
function for a free Fermi-gas of quarks [solid line; see also
Eq.~(\ref{rho2fg})]. Evidently, Pauli correlations play an
insignificant role in this very dilute system.  As the density
increases to $\rho\!=\!0.6$, $f_{s}$ remains at zero but nucleon
swelling --- the increase in the length-scale for confinement --- is
now discernible as indicated by the deviation from the isolated-nucleon
result. Moreover, the correlation function now displays the
typical long-distance oscillations, albeit much stronger, of a
Fermi-gas system. In the third panel ($\rho=1$) the system displays
features related to both --- hadronic and Fermi-gas --- limits: an
enhanced correlation at short distances and a rapid ``healing''
towards the Fermi-gas limit at larger distances. Finally, the last 
panel shows that the evolution to the Fermi-gas limit is now 
complete, in spite of a value for the variational parameter which, 
at 50\% of its isolated-hadron value, is still large.

\begin{figure}
\bigskip
\centerline{
  \psfig{figure=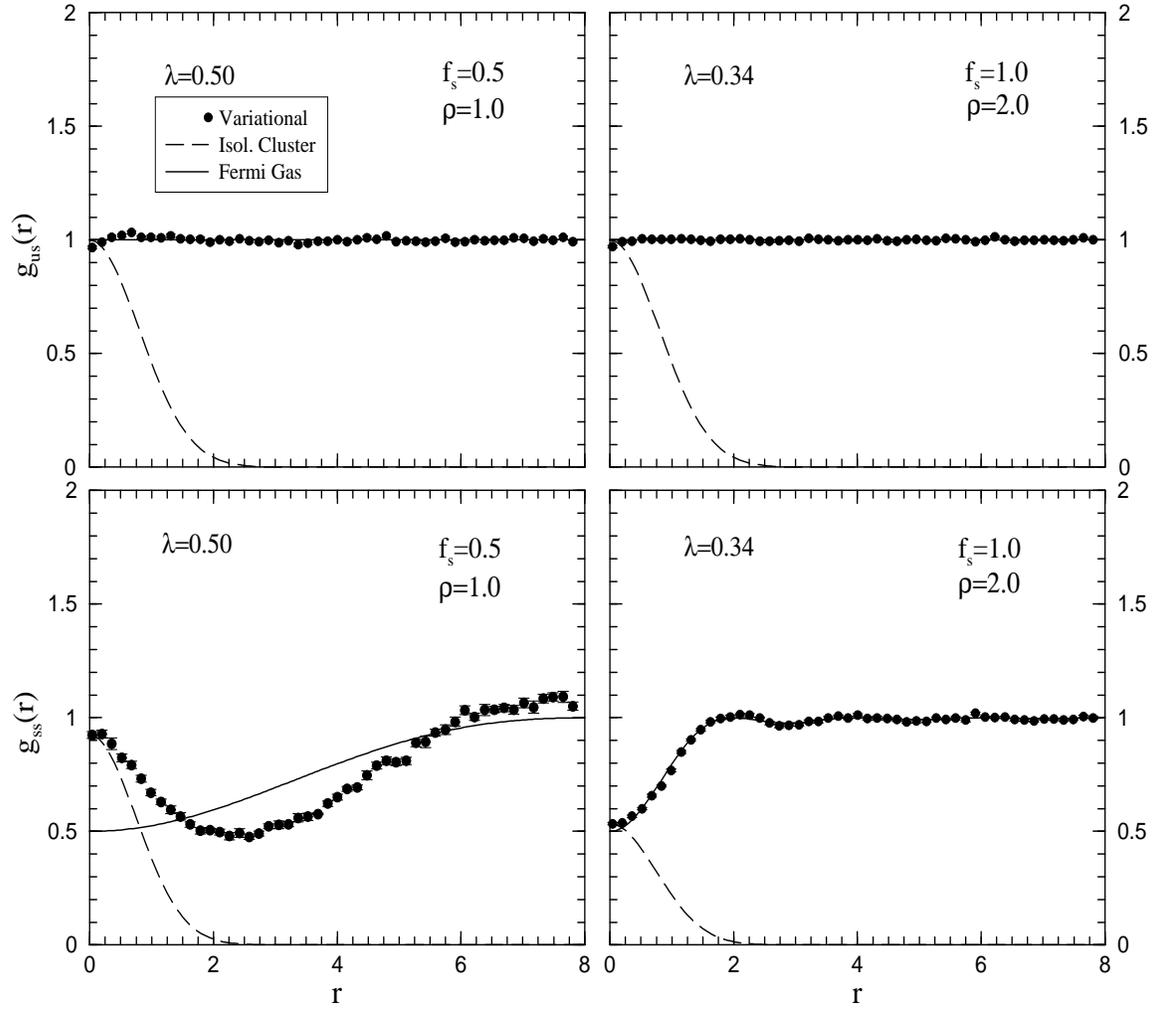,height=5.5in,width=6in,angle=-90}}
 \vskip 0.1in
 \caption{Two-body up-strange and strange-strange correlation 
          functions for a variety of densities.}   
 \label{fig6}
\end{figure}

In addition to the up-up correlation function one can measure two-body
correlations for the other two flavor combinations as displayed in
Fig.~\ref{fig6}. The upper two panels show, as a result
of the Pauli principle not being operative between quarks of different
flavors, a featureless ``up-strange'' correlation function. In
contrast, the strange-strange correlation function at intermediate
density (lower-left panel) shows characteristics that cannot be 
associated exclusively with either those of isolated clusters or a 
Fermi gas but contains some features of both.  
At the largest density shown in the figure ($\rho\!=\!2$) the
transition to the Fermi-gas limit seems now complete.

\begin{figure}
\bigskip
\centerline{
  \psfig{figure=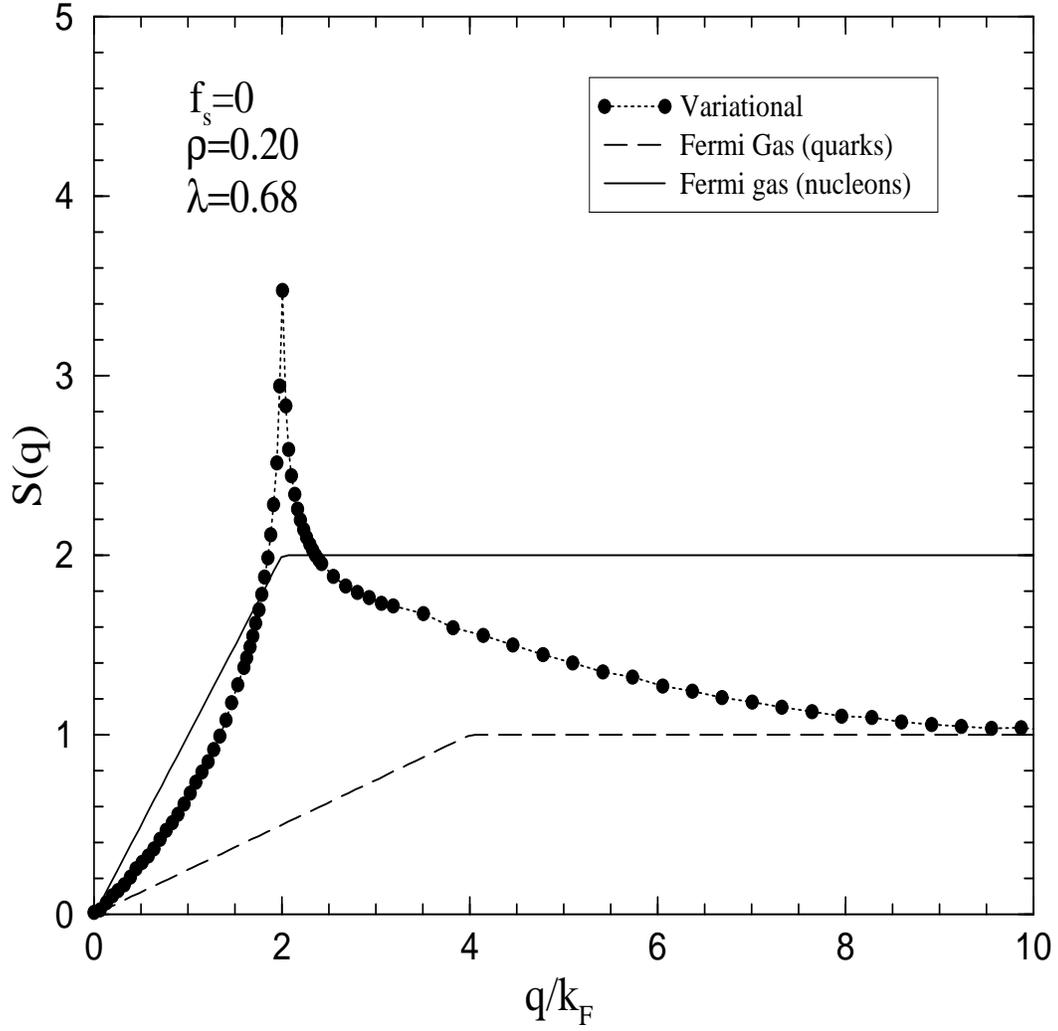,height=5.5in,width=5.5in,angle=0}}
 \vskip 0.1in
 \caption{Coulomb sum as a function of density in units of 
	  the Fermi momentum.}
 \label{fig7}
\end{figure}

So why the emphasis on the two-body correlation function? As stated
earlier, the two-body correlation function is a useful observable for
characterizing the transition from hadronic to quark matter. Moreover,
it displays clearly the dynamic interplay between clustering and Pauli
correlations. Yet its real appeal stems from its relation to the dynamic 
response of the system. Indeed, the two-body correlation function is 
related to the Fourier transform of the static structure factor:
\begin{equation}
   \frac{1}{N}S(q)=1+\rho\int_{-\infty}^{\infty}
                     e^{-iqr}\Big(g(r)-1\Big)dr
                  =1+\rho\int_{0}^{\infty}
                   2\cos(qr)\Big(g(r)-1\Big)dr \;,
 \label{Fourier}
\end{equation}
with the static structure factor, itself, being defined as the
integrated sum of the dynamic response function,
\begin{equation}
   S(q)=\int_{0}^{\infty}S(q,\omega)\,d\omega \;.
 \label{sofq}
\end{equation}
As a preview of future work, devoted to the study of the exact
Euclidean response of the system and its analytic continuation 
into real time, we focus on the static structure factor known 
as the Coulomb sum. In Fig.~\ref{fig7} we display the Coulomb 
sum as a function of the momentum transfer at the low density 
of $\rho=0.2$. The dashed line represents the response of free 
Fermi gas of quarks; evidently well below the variational result. 
This behavior suggests that at these momentum transfers both 
quarks in a hadron respond coherently. Yet the coherence is 
incomplete, as the response of the system at low $q$ is below 
that of a free Fermi gas of nucleons (solid line). Some of the
coherence is lost because of the intrinsic quark structure of the
nucleon: the nucleon ``survival'' probability is proportional to 
its elastic form factor. Note that the very large ``spike'' at 
exactly $q=2k_{F}$ is an artifact of the ``rigidity'' imposed 
by the Pauli exclusion principle in one spatial dimension. We are 
currently in the process of applying familiar nuclear techniques, 
such as the impulse approximation, to gain insight into the 
physics of this fundamental problem. Preliminary results for a 
simplified version of the model have been published in
Ref.~\cite{Pie97}.

\section{Conclusions and Outlook}
\label{sec:conclusions}
We have employed variational Monte Carlo techniques to compute
ground-state properties of strange matter in a one-dimensional 
string-flip model with two flavors and two colors of quarks. In 
this simplified model baryons are color-neutral, two-quark 
composites. We have used a many-body potential that confines 
quarks within baryons yet enables the hadrons to separate without 
generating unphysical long-range forces as the confining force 
saturates within each individual hadron. The many-body nature of 
the potential entails a difficult optimization problem: how to pair 
$N$ quarks into $A\!=\!N/2$ color-neutral clusters so that the 
overall flux-tube energy is minimized. For the general 
three-dimensional case this problem involves searching among $A!$ 
distinct configurations, a task that becomes prohibitely costly 
even for small systems. The one-dimensional case studied here is 
special as one can restrict the search to only $A$ configurations. 
In this manner we were able to simulate systems with up to $N=80$ 
quarks (note that in an exhaustive routine one would have to search 
among close to $10^{48}$ configurations!). In this paper we have 
used, for the sake of computational expediency, the simplest 
string-flip model that can describe the basic quark structure of 
strange matter.

The crucial feature of all string-flip models is the need to determine
an optimal grouping of quarks into hadrons. This need alone yields two
natural limits for the model: a hadronic limit at low density and a
Fermi-gas limit at high density. The system thus exhibits a transition
from hadronic matter, where clustering correlations dominate, to quark
matter, where only Pauli correlations remain.  This is one of the
greatest advantage of the model: the transition from hadronic- to
quark-matter is dynamic, as there is no need to resort to ad-hoc
prescriptions. The main goal of this paper was the characterization of
the transition to strange matter; does the transition happens at a
density at which the system has already dissolved into a collection of
``free'' quarks or do clustering correlations remain strong? In our
model the transition to strange matter occurred at a density of
approximately $\rho\!=\!1.3\rho_{c}$ ($\rho_{c}$ was defined as the
transition density predicted by the free Fermi-gas estimate). Thus,
the many-body potential delays the transition relative to that of a
free Fermi gas. More interestingly, the length-scale for confinement
increased at the transition by merely 10\% of its isolated-hadron
value and in fact clustering correlations remained strong well into the
transition region.  Indeed, we observed that ground-state properties
computed in the ``hypernuclear'' limit were almost identical to those
predicted by the variational approach. In sharp contrast, a
Hartree-Fock estimate --- which includes Pauli but not clustering
correlations --- predicted a much larger value for the transition
density and an energy-per-quark that overestimates (at the transition
density) the variational estimate by about 20\%. Thus --- on the basis
of our findings --- we conclude that the behavior of the system at the
transition density is well reproduced by a hadronic model with no
baryon swelling, but not by one that ignores clustering correlations.

In the future we plan to extend this calculation in several
directions. First, we need to refine the many-body potential in order
to study the possibility of stable, or absolutely stable, strange
matter. The string-flip potential used here seems to account for the
short- and intermediate-range structure of the nucleon-nucleon
potential. Indeed, the Pauli principle between quarks generates a
short-range repulsion between clusters while quark-exchange generates,
at least part of, the intermediate-range attraction. What is missing
from the model is the long-range (pion) tail. One could introduce a
phenomenological long-range part to the potential constrained to
reproduce the equation of state of nuclear matter. With this model for
the potential in hand one could then predict the stability, or
lack-thereof, of strange matter. Second, while undoubtedly
challenging, simulating three-dimensional strange matter with three
flavors and three color of quarks is now within computational
reach. Finally, one should attempt to calculate the dynamic response
of the system. We wish to understand --- as a function of density and
of the momentum transfer --- when do leptons scatter from hadrons and
when do they scatter from the individual quarks. We have offered a
preliminary answer to this question. By computing the two-body
correlation function, and from it the Coulomb sum, we observed that
the response of the system at small momentum transfers was well above
the response of a free Fermi gas of quarks. This suggested that all
the quarks in the hadron respond coherently. Yet we saw that the
coherence was incomplete, as the response was below that of a Fermi
gas of nucleons. We attributed this loss of coherence to the internal
quark structure of the hadron. How this picture can be reconciled
with more sophisticated models of the reaction, such as in the impulse
approximation, remains an interesting open question. In summary, we
have shown that in spite of their deceiving simplicity, string-flip
models of hadronic matter display rich behavior that could yield
valuable insights into the physics of strange matter.

\acknowledgments
This work was supported by the DOE under Contracts 
Nos. DE-FC05-85ER250000 and DE-FG05-92ER40750.

\end{document}